# Link and Location Based Routing Mechanism for Energy Efficiency in Wireless Sensor Networks

Ms.Pavithra.G[1], Ms.Devaki.P[2]

*1 – Student, Department of CSE, Kumaraguru College of Technology, Coimbatore-49*
*2 – Associate Professor, Department of CSE, Kumaraguru College of Technology, Coimbatore-49*
*Tamil Nadu, India.*

*Abstract*— **In Wireless Sensor Networks, sensed data are reported to the sink by the available nodes in the communication range. The sensed data should be reported to the sink with the frequency expected by the sink. In order to have a communication between source and sink, Link based routing is used. Link based routing aims to achieve an energy efficient and reliable routing path. This mechanism considers the status (current energy level in terms of Joules) of each node, link condition (number of transmissions that the Cluster Head (CH) and the Gateway (GW) candidates conducts) and the transmit power (power required for transmission in terms of Joules). A metric called Predicted Transmission Count (PTX) for each node is calculated using its status, link condition and transmit power. The node which has highest PTX will have the highest priority and it will be the potential candidate to act as CH or GW. Thus the selection of proper CH or GW reduces the energy consumption, and the network lifetime is increased.**

*Keywords*——**Wireless sensor network, energy efficient, clustering, link based, routing.**

## I. INTRODUCTION

The Wireless sensor network (WSN) is promising network architecture. It used in many applications such as monitoring the environment, detecting the object, tracking the event, and security surveillance [1]. WSNs consist of large numbers of very small autonomous wireless devices, called sensor nodes. Sensor nodes perform multiple functions such as sensing, computing, and communication [2]. In typical WSNs, a central node called sink, which sends query message to other nodes in the network. The sensor nodes (i.e., source nodes) must report the sensing or monitoring data to the sink. Charging batteries for sensor nodes is often difficult since the sensor nodes are battery-powered devices. Data transmission is the major source of energy consumption compared to sensing, communication, and computation [3]. Thus, it is important to design an energy efficient routing scheme for reporting the sensed data to achieve a high delivery ratio and to increase the lifetime of network.

The three categories of routing protocols available in the existing systems of WSNs: chain-based, tree-based, and cluster-based [4]. In the chain-based routing protocol, the farthest node from the sink becomes the leader. It uses more energy to transmit messages to the sink. The chain length increases when the number of sensor nodes on the chain increases. Thus a transmission delay is generated and additional energy is consumed because of multi-hop transmission. In the tree-based routing protocol, the node present at the root becomes the leader. Bottleneck occurs at the root node and hence it exhausts its battery power quickly. In addition, the protocol generates higher transmission delay when the tree has more levels. Clustering is introduced to support route determination in WSNs [3], [5]–[7]. Clustering groups all the sensor nodes into multiple clusters. In each cluster, one node is elected as the clusterhead, which controls the cluster. Gateways are used to connect multiple clusters. Clustering is effective in one-to-many, many-to-one, one-to-any, or one-to-all communications, and it can improve the routing performance [8], [9].

In clustering, it is very difficult to select proper nodes to act as clusterheads and gateways. Many clusterhead election approaches are proposed for construct-ing clusters [3], [8], [10], [11]. Each node locally exchanges messages with the nodes in its communication range (i.e., neighbors nodes) to determine whether it should become a clusterhead. These approaches are known as active clustering. Kwon and Gerla [12] have proposed a clustering technique, called passive clustering. In the passive clustering technique, all the nodes in a cluster have an external cluster state. The clusterhead and gateway nodes are major participants in packet delivery. When a node receives a data packet, it determines whether it must change its current state depending on its current state and state of the sender of the packet. Each node piggybacks its state onto the transmitted packet. Thus a node can realize the cluster states of all its neighbors. The passive clustering technique reduces the communication overhead by effectively decreasing the number of explicit control packets to constantly maintain the cluster information. Existing clustering approaches use a random strategy to determine clusterheads if there are many clusterhead candidates. Sometimes this strategy determines improper clusterheads. Generally clusterhead consumes more battery power than other nodes. The routing path may be destroyed when the clusterhead exhausts its battery power. The packet delivery ratio is reduced. If the clusterhead is associated with a poor quality link, it consumes energy unnecessarily since it causes additional retransmissions.

A link-based routing mechanism is proposed based on the passive clustering technique to support energy-efficient routing in WSNs. This technique determines proper nodes to become clusterheads and gateways. Thus it provides persistent and reliable routing path. In the link-based routing, the node status (e.g., residual energy) and link condition (e.g., quality) are used by the clusterhead and gateway candidates to determine a clustering metric. The metric is called the predicted trans-mission count. The number of transmissions that clusterhead and gateway candidates conduct is called the predicted transmission count. Consumption of transmit power, residual energy and link quality are used to measure the metric. The predicted transmission count derives the priority for the clusterhead and gateway candidates to evaluate its qualification for a clusterhead or a gateway. The candidate having the highest priority is elected as a clusterhead or a gateway. Thus the link-based routing supports energy-efficient routing.

The rest of this paper is organized as follows. Section II has the introduction about the clustering metrics and the traditional passive clustering technique. Section III describes the network model and assumptions. Section IV presents the existing link-based routing in detail. Section V provides the simulation results. Section VI describes the proposed ides in brief. Section VII presents the conclusion.

## II. PRELIMINARIES

This section explains the concept of clustering metrics and presents the original passive clustering technique.

### A. Clustering Metrics

Many metrics have been proposed for clusterhead election to construct an efficient cluster structure. The metrics include node identifier (ID) [8], node degree [10], [12], [13], and residual energy [3], [14]. In general, a cluster consists of only one clusterhead, and each node has a unique ID. Thus the node ID is considered as the metric for clusterhead election. A node having the largest or smallest ID value among nodes in its proximity is elected as a clusterhead. If the ID of the selected clusterhead is smaller than that of the other cluster members then it is called The Lowest ID Cluster (LIC) algorithm [8].

Previous studies have proposed an alternative clusterhead election strategy based on the node degree. The node degree indicates the number of nodes in a node's communication range [10], [12], [13]. To measure the node degree, each node periodically exchanges messages with all the one-hop count neighbors, and counts the number of messages received back. The random-based clusterhead election approach selects a clusterhead that has no neighbors. Since this approach has no neighbors it fails to construct a cluster structure and discover a routing path. If the node with smaller node degree value is selected to become a clusterhead, it may create a large number of clusters. Thus a considerable amount of overhead for cluster maintenance is generated and delivery delay is increased. Therefore, a Highest Connectivity Clustering (HCC) algorithm is proposed by Gerla and Tsai [10], where a node is selected as the clusterhead if its degree (i.e., connectivity) is larger than that of all its neighbors.

Sensor nodes are energy-constrained, and selecting a node with more residual energy as the clusterhead is a major challenge in clustering [3], [14]. A cost function to evaluate the qualification of a node for a clusterhead is used in the Hybrid Energy-Efficient Distributed (HEED) Clustering Approach [3]. Assume that all the nodes are initially fully charged. The probability of becoming a clusterhead, $CH_{prob}$, is determined using the cost function. It is derived as

$$CH_{prob} = C_{prob} \times \frac{E_{res}}{E_{ini}} \qquad (1)$$

where $C_{prob}$ is the percentage of clusterheads that are present initially in the network. $E_{res}$ is the estimated current residual energy. $E_{ini}$ is the initial energy (i.e., maximum energy) of a node. The node derives higher $CH_{prob}$ if it has high residual energy. Thus the node with the maximum amount of residual energy in its proximity will successfully become a clusterhead.

### B. Passive Clustering (PC) Technique

Kwon and Gerla proposed a passive clustering (PC) technique for construction of a cluster structure [12]. Instead of extra explicit control packets, the PC uses on-going data packets. Thus the PC can reduce the control overhead during construction and maintenance of clusters. To represent a node's role in a cluster the PC technique uses five external states. Each node possesses an external state. The external states are initial node (IN), ordinary node (OD), clusterhead node (CH), gateway node (GW), and distributed gateway node (D_GW). To represent the tentative role of a node the PC technique also introduces two internal states, clusterhead ready node (CH_R) and gateway ready node (GW_R). If a node in the external state receives data packets, it may change its current state. When a node sends out a data packet then the node in the internal state must enter the external state.

The two innovative mechanisms are proposed by PC technique: First Declaration Wins mechanism and Gateway Selection Heuristic mechanism. These two mechanisms are used to determine CH and GW nodes. The First Declaration Wins mechanism has a contention strategy which is used by the CH candidate (i.e., CH_R) to declare that it wants to become a CH node. The clusterhead candidate first claiming to become a CH node within the communication range will successfully become a clusterhead node. The Gateway Selection Heuristic mechanism determines the minimal number of GW nodes. It guarantees that a single cluster has at least two GW nodes to maintain network connectivity.

In PC, clusterhead and gateway nodes dominate the energy usage since they are the main participants in data transmission. Random selection strategy is used by the PC technique to determine CH and GW nodes. Although the PC technique is an effortless approach, it does not consider the node status and link condition in clustering. Hence it is not an efficient approach.

### III. NETWORK MODEL AND ASSUMPTIONS

This study considers the network which is an undirected graph $G = (V, E)$, where $V$ is the set of nodes and $E \subseteq V \times V$ is the set of links between two neighboring nodes. Let the link between two nodes, $s_i$ and $s_j$ is denoted by $e_{ij} \in E$. Let $\mathbf{S}^{nbr}{}_i$ be the set of links between the neighboring nodes, and $N_i^{nbr}$ be the number of elements in $\mathbf{S}^{nbr}{}_i$. Let the current and new cluster states of node $s_i$, is denoted by $ST_i^{cur}$ and $ST_i^{new}$ respectively. The cluster identifier of node $s_i$ is $ID(i)$. The sink sends query messages to the nodes (i.e., source nodes) in a specific area of interest. The sensing data is periodically reported to the sink by the source nodes. This report should satisfy the quality that the sink expects. This study considers the report quality as that the reporting frequency must exceed a pre-defined threshold, denoted as $N_{req}$. $N_{req}$ is carried in query messages. Let $q_{ij}$ is the predicted transmission count of $e_{ij}$ and $\rho_i$ denotes the priority of candidate $s_i$.

This study assumes that all sensor nodes are stationary and have the same communication range. And also each sensor node has a unique identifier and is equipped with a Global Positioning System device to measure its physical location. To ascertain the number of neighbors the nodes periodically exchange a message. Its assumed that the source node knows the reporting frequency ($N_{req}$) that the sink requires. Assume that the report messages are fixed in size. Also assume that all nodes have the same electronics energy for a unit of data and the same amplifier energy for transmitting a unit of data over a unit distance.

### IV. PROPOSED LINK-AWARE CLUSTERING MECHANISM

This section deals with the proposed predicted transmission count and the procedure of priority calculation in the proposed link-based routing mechanism, followed by an example of routing operation.

#### A. Predicted Transmission Count

Random selection is an effortless strategy to determine CH and GW nodes but it is not an efficient approach because of its disregard of node status and link condition. Using only a single factor cannot expose the influence of other factors on routing performance. The link-based routing mechanism considers node status and link condition. Based on this it proposes a novel metric, called the predicted transmission count (PTX). The PTX evaluates the suitability of CH or GW candidates and it represents the capability of a candidate for persistent transmission to a specific neighboring node. The transmit power, residual energy, and link quality is used by the link-based routing mechanism to derive the PTX of CH or GW candidate. A node having a large PTX value indicates a high likelihood of becoming a CH or GW node.

The link reliability often depends on the channel condition since the channel condition of wireless links varies with time. If a node is associated with an unreliable link, the data delivery is likely to fail. Hence it leads to packet retransmissions. Thus the candidate node associated with a stable link is preferred to be selected as a CH node or a GW node. The expected transmission count, called ETX is used by the previous researches, to evaluate the level of link quality [15], [16]. The proposed link-based routing mechanism also uses the ETX to measure the expected bi-directional transmission count of a link. Let the ETX of link $e_{ij}$ is $ETX_{ij}$, and therefore $ETX_{ij}$ can be defined as

$$ETX_{ij} = \frac{1}{p_{ij}^f \cdot p_{ij}^r} \qquad (2)$$

where $p_{ij}^f$ denotes the forward delivery ratio and $p_{ij}^r$ denotes the reverse delivery ratio from node $s_i$ to node $s_j$. The measured probability that indicates that a data packet successfully arrives at the recipient is called forward delivery ratio. The probability that indicates that the acknowledgment (ACK) packet is successfully received is called reverse delivery ratio. Each node in the link-based routing mechanism periodically broadcasts a message to find the distance between the nodes, forward delivery ratio, and reverse delivery ratio of its neighbors. Hence it is possible to determine the ETX. When node $s_i$, receives the report messages from the node $s_j$, it can use Eq. (3) to derive the PTX, $q_{ij}$

$$q_{ij} = \frac{E_i^{res}}{ETX_{ij} \cdot E^{tx}(k, d_{ij})} \qquad (3)$$

where $E_i^{res}$ is the residual energy of the node $s_i$, $d_{ij}$ is the distance between two nodes $s_i$ and $s_j$, and $E^{tx}(k, d_{ij})$ is the energy consumption for the node $s_i$ to transmit a $k$-bit message over the distance $d_{ij}$.

This study considers the first order model for the radio hardware energy dissipation [17]. Transmitters dissipate energy to run the radio electronics and the power amplifier. Let $E_{elec}^{tx}(k)$ denotes the energy consumption of the radio electronics and $E_{amp}^{tx}(k, d_{ij}^n)$ denotes the power required by the amplifier, to transmit a $k$-bit message over a distance $d_{ij}$. The total energy consumed, $E^{tx}(k, d_{ij})$ can be derived from

$$E^{tx}(k, d_{ij}) = E_{elec}^{tx}(k) + E_{amp}^{tx}(k, d_{ij}^n) \qquad (4)$$

In link-based routing mechanism, the first order model uses both the free space and multipath fading channel models [18]. The free space model is adopted when the distance between the transmitter and the receiver is less than a pre-defined threshold, denoted as $d_0$. Otherwise, the multipath model is adopted. Let $E_{elec}$ denotes the electronics energy. Electronics energy is related to the modulation, digital coding and filtering techniques.

$E^{tx}(k, d_{ij})$ in Eq. (4) can be written as

$$E^{tx}(k, d_{ij}) = \begin{cases} k.E_{elec} + k.\epsilon_{fs}.d_{ij}^2, & d_{ij} < d_0 \\ k.E_{elec} + k.\epsilon_{mp}.d_{ij}^4, & d_{ij} \geq d_0 \end{cases} \qquad (5)$$

where $\epsilon_{fs}\cdot d_{i\,j}^2$ or $\epsilon_{mp}\cdot d_{i\,j}^4$ is the amplifier energy. The amplifier energy is related to the distance between the transmitter and the receiver and the acceptable bit error rate.

### B. Priority Calculation

The link-based routing mechanism evaluates the suitability of CH or GW candidates to determine proper participants to forward data packets. A clusterhead candidate (CH_R node) or a gateway candidate (GW_R node), $s_i$ , performs the following steps to determine its priority.

*Step 1:* Calculate the PTX value of each neighboring.

*Step 2:* Divide the neighbor nodes $\mathbf{S}^{nbr}_i$ into two subsets, $\mathbf{S}_{sat}$ $(i)$ and $\bar{\mathbf{s}}_{sat}(i)$, where the PTXs of all elements in $\mathbf{S}_{sat}(i)$ are greater than or equal to the threshold value denoted by $N_{req}$ , and the PTXs of all elements in $\bar{\mathbf{s}}_{sat}(i)$ are smaller than $N_{req}$.

*Step 3:* If $\mathbf{S}_{sat}\ (i\ ) \neq \emptyset$, set the priority, $\rho_i$ as the PTX of the node, which has the minimum PTX in $\mathbf{S}_{sat}\ (i\ )$; otherwise, set $\rho_i$ as the PTX of the node, which has the maximum PTX in $\bar{\mathbf{s}}_{sat}$ $(i\ )$.

Based on the definition of the PTX, a candidate node which derives a large PTX value if it connects to nodes with a higher quality or if it supports more transmission counts. The proposed link-based routing mechanism determines the candidates satisfying the report quality by putting them into the subset $\mathbf{S}_{sat}\ (i\ )$. If the subset $\mathbf{S}_{sat}(i)= \emptyset$, the link-based routing mechanism considers the minimum PTX of all PTXs as the priority of the node $s_i$ since the link corresponding to the minimum PTX can adequately support the report quality. If no link is able to satisfy the report quality (i.e., $\mathbf{S}_{sat}\ (i\ ) = \emptyset$), this study selects the link that can support as many message reports as possible. Thus, the link-based routing mechanism considers the maximum PTX of all PTXs in $\mathbf{S}_{sat}\ (i\ )$ as the priority of the node $s_i$ . Algorithm 1 shows the detailed procedure of priority calculation of candidate nodes in the link-based routing mechanism.

To make sure that the high priority node becomes the CH or GW node, the link-based routing mechanism uses a random backoff approach to defer the transmission of data packets. Let the waiting period of candidate node $s_i$ be $T_i^w$. Then, $T_i^w$ can be derived from

$$T_i^w = t_{slot}\cdot \theta\left(\frac{1}{\rho_i}\right) \qquad (6)$$

where $t_{slot}$ is the unit of time slot, and $\Theta(x)$ rounds the value of $x$ to the nearest integer less than or equal to the value of $x$ .

### C. Cluster State Transition

The cluster state transition diagram of the proposed link-based routing mechanism is shown in the Figure 1. Algorithm 2 is used by a node when receiving messages to determine whether it must change its current state. This paper uses the IN node as an example to explain the state transition of link-based routing mechanism (i.e., from line 2 to line 11 of Algorithm 2). If an IN node receives messages from either a CH node or a GW node, it changes its cluster identifier as that of the sender. This is because of this IN node and the sender belongs to the same cluster.

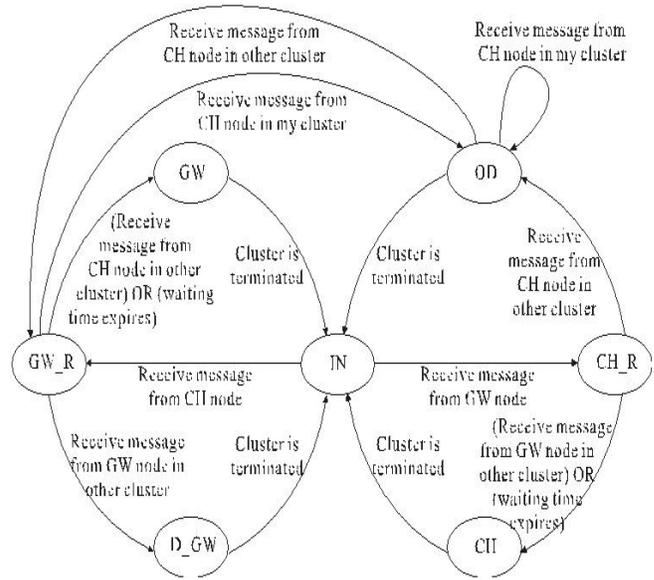

Fig.,1. Cluster state transition diagram of the proposed link-based routing mechanism.

**Algorithm 1** Cluster State Transition Algorithm in the Link-based routing mechanism

1 /* To be performed by node $s_i$ when it receives report messages from node $s_j$ . /*

    **Input:** $ST_i^{cur}$, $ST_j^{cur}$ , $ID(j)$.

    **Output:** $ST_i^{cur}$ .

2 **switch** $ST_i^{cur}$ **do**

3     **case** *IN*

4         $ID(i) \leftarrow ID(j)$;

5         **if** $ST_j^{cur} = CH$ **then**

6             $ST_i^{cur} \leftarrow GW\_R$;

7             **call Procedure** Contention;

8         **else**

9             **if** $ST_j^{cur} = GW$ **then**

10                 $ST_i^{cur} \leftarrow CH\_R$;

11                 **call Procedure** Contention;

12     **case** *OD*

13         **if** $(ST_j^{cur} = CH)$ **and** $(ID(i)\_= ID(j))$ **then**

14             $ST_i^{cur} \leftarrow GW\_R$;

15             **call Procedure** Contention;

16         **else**

17             $ST_i^{new} \leftarrow ST_i^{cur}$ ;

18     **otherwise**

19         $ST_i^{new} \leftarrow ST_i^{cur}$ ;

If the sending node is a CH node, the IN node then transits its state to GW_R state. Otherwise, if the sending node is a GW node, then the IN node transits its state to CH_R. Meanwhile, the IN node enters the contention procedure (i.e., Algorithm 3) to calculate its priority and determine its current state. The node forwards the received message if it becomes a CH or GW node.

---

**Algorithm 2** Procedure Contention

---

1 Calculate $\rho_i$ ; /* Algorithm 1
2 Determine $T_i^w$ ; /* (6)
3 $i\ s\ New\ State\ Deter\ mi\ ned \leftarrow 0;$
4 **while** $T_i^w$ does not expire **do**
5    **if** *receive a report message from* $s_k$ **then**
6      **if** $ID(i)\_=ID(k)$ **then**
7        **if** $ST_k^{cur}=CH$ **then**
8          if $ST_i^{cur}=GW\_R$ then
9          $ST_i^{new} \leftarrow OD;$
10        **Else**
11          $ST_i^{new} \leftarrow GW;$
12      **else if** $ST_k^{cur}=GW$ **then**
13        **if** $ST_i^{cur}=CH\_R$ **then**
14          $ST_i^{new} \leftarrow CH;$
15        **else**
16          $ST_i^{new} \leftarrow D\_GW;$
17      $i\ s\ New\ State\ Deter\ mi\ ned \leftarrow 1;$
18 **if** $i\ s\ New\ State\ Deter\ mi\ ned = 0$ **then**
19    **if** $ST_i^{cur}=GW\_R$ **then**
20      **if** *receive no report message from the CH*
     *neighbors during* $T_i^w$ **then**
21        $ST_i^{new} \leftarrow GW;$
22    **else**
23      $ST_i^{new} \leftarrow OD;$
24 $ST_i^{new} \leftarrow ST_i^{cur};$
25 **return** $ST_i^{new}$

---

## V. SIMULATION RESULTS

This study used ns-2 as the network simulator and conducted numerous simulations with different number of sensor nodes to evaluate the performance. This study evaluates the following main performance metrics: Energy consumption and the number of Active Nodes.

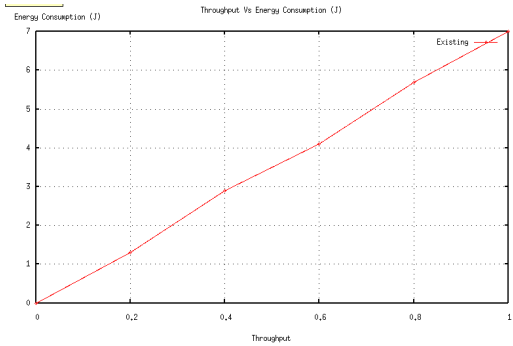

Fig. 2 Throughput - Energy consumption

Figure 2 shows the energy consumed by the sensor nodes with respect to the throughput which is obtained from the system.

Similarly Figure 3 shows the number of active nodes when the simulation time keeps on increasing.

By the process of selecting the candidate with the highest priority as a CH or a GW, the system can guarantee that the discovered routing path can remain persistent.

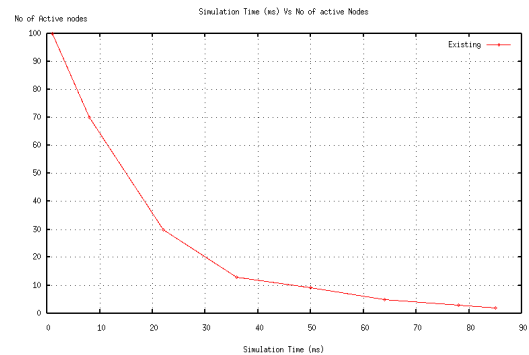

Fig.3.Simulatiom Time - No.of Active Nodes

## VI. PROPOSED WORK

The existing link based routing mechanism, considers both link condition (i.e., ETX value) and node status (i.e., energy usage). It can efficiently construct a reliable and persistent routing path to guarantee the report quality. The link quality is calculated by finding the distance of each neighboring node throughout the routing path. It consumes comparatively more energy when each of the neighboring nodes is taken into consideration. To overcome the above problem, the energy efficient GPSR protocol is used.

Greedy Perimeter Stateless Routing is a location-based routing protocol. In GPSR, each node should be capable of determining its own location and the source will be aware of the location of destination. Source includes the location of destination in the header of every packet. If the destination is not directly reachable, the source starts with greedy forwarding i.e source node forwards the data packet to the neighbor node that is closest to the destination in the coordinate space. Such greedy forwarding is repeated at every intermediate node until the destination is reached.

## VII.  CONCLUSION

The location-based routing protocol improves the energy efficiency when compared with the link-based routing protocol. The location-based routing protocol considers the energy level and geographical information to route a packet towards the target region. It maintains only the location information rather than maintaining the routing information. Gateway nodes are not needed since clusterhead and cluster members can be used as intermediate nodes. Thus it saves more energy and increase the network lifetime than the link based routing protocol.